# Emotionally-Informed Decisions: Bringing Gut's Feelings into Self-adaptive and Co-adaptive Software Systems

Emmanuelle Tognoli
Center for Complex Systems & Brain Sciences
Florida Atlantic University
Boca Raton, FL 33431, USA
etognoli@fau.edu

Shihong Huang
Computer & Electrical Engineering and Computer Science
Florida Atlantic University
Boca Raton, FL 33431, USA
shihong@fau.edu

## ABSTRACT

Software systems now complement an incredibly vast number of human activities, and much effort has been deployed to make them quasi-autonomous with the build-up of increasingly performant self-adaptive capabilities, so that the burden of failure, interruption and functional loss requiring expert intervention is fewer and far in between. Even as software systems are rapidly gaining skills that beat humans', humans retain greatly superior adaptability, especially in the context of emotionally-informed decisions and decisions under uncertainty; that is to say, self-adaptive and co-adaptive software systems have yet to acquire a "gut's feeling". This provides the double opportunity to conceptualize human-inspired processes of decision-making under uncertainty in the self-adaptive part of a software, as well as to source human unique emotional competences in co-adaptive architectures. In this paper, some algorithms are discussed that can provide software systems with realistic decision-making, and some architectures are conceptualized that resort to human emotions to quantify uncertainty and to contribute in the software's adaptation process.

## 1 INTRODUCTION

In Plato's Cratylus, Heraclitus of Ephesus was cast as saying "Everything flows and nothing abides". Likewise, software systems are deployed in an every-changing environment and their adaptation to uncertainty and change has become a matter of intense research. In the present review, we examine how human-inspired decision-making processes in software systems and emotional inputs from Humans-in-the-Loop have the potential to improve software systems' adaptation to uncertainty and change. Decision making is biology's response to uncertainty (Hampshire & Hart, 1958; Hastie & Dawes, 2001; Gold & Shalden, 2007; Doya, 2008). That is, faced with the fact that contingencies between action and outcome are difficult to predict, organisms have evolved some processes that enhance action selection and ultimately support survival. Furthermore, emotions, for long seen as a passion detrimental to reason, have emerged as a positive contributor to sound decision-making (Elster, 1991; Schwarz, 2000; Hastie & Dawes, 2001; Lowenstein & Lerner, 2003; Bechara, 2004; Naqvi et al, 2006). In this paper, we review the psychophysiological literature of decision-making (section 2) and emotion (section 3), before conceptualizing an adaptive system (section 4) employing continuously measured emotions (section 5) to perform autonomous and joint tasks (section 6).

## 2 A BIRDEYE VIEW ON DECISION-MAKING

Decision making is an enormous and interdisciplinary field of inquiry with dispersed terminology, conceptual (nomological) network, paradigms and perspectives. In the following, we outline the basic operation of this cognitive function. Whenever possible, and sometimes at the cost of oversimplification, we try to bring together closely-related concepts so that readers can relate the overview with a broader range of its associated literature. Then, we discuss the interplay of decision-making with a wide array of neurocognitive functions, and we identify key modulators that affects the process: uncertainty, risk, time and emotions.

### 2.1 A theoretical model of decision-making

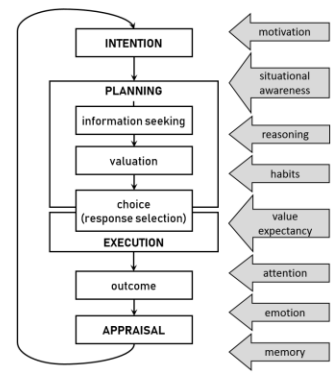

Figure 1: the processes involve in decision making (left) and the neurocognitive functions that support them (right).

The core element of all theories of decision-making is the notion of choice between alternatives (Fig.1 center left, (Hogarth, 1987; Vershure et al., 2014) and it is foundational to the hypothesis that living organisms, especially advanced ones like humans, differ from inanimate matter (including technological artifacts) in that they are supposedly endowed with free-will (but see Libet, 1999; Soon et al., 2008. The motivational driving force of behavior, what sets the organism in motion in the first place and triggers the cascade of cognitive processes involved in decision-making, is an intention or goal (Fig.1 top left (Hampshire & Hart, 1958; Bratman, 1987). An intention engages a process of planning (Fig.1 left, Azjen, 1991), which can be conscious or unconscious (Soon et al., 2008); and fast and automatic or slow and deliberative (Norman & Shallice, 1980; Kahneman, 2011). Planning requires the acquisition of information about the context in which the behavior is to take place (Endsley, 1995), both with sensing/monitoring the environment, and within the agent him/herself (internal states, abilities, disposition, and priors). In a complex environment, planning eventually runs into the multiple paths available to achieve said goal. At each of those bifurcations, a choice must be made between the different alternatives. A key factor presiding in the choice is a series of mental operations that predict the value (utility, fit) of different alternatives (Fig.1 left, Higgins, 2000; Shidara & Richmond, 2002; Sugrue et al., 2005; Gold & Shalden, 2007; Hare et al., 2009), so that the organism choses in his/her best interest. Once a choice has been executed, the organism's actions precipitate an outcome (Fig.1 left), whose actual value is often returned to the organism in the form of a feedback (Saati, 1996): reward when positive, punishment when negative. The feedback loop sets internal processes of appraisal (a comparison between expected and actual outcome, Gross, 2002; Shihara and Richmond, 2002) that engages the emotional system: for instance, when the actual value of a choice falls short of the expected value, human will experience regret (Zeelengberg, 1999). And this appraisal process contributes to future decisions made in similar contexts (Fig.1 bottom left), via reinforcement learning (Doya, 2008; Curtis & Lee, 2010; Rushworth et al., 2011). At the next iteration of decision making, this accrued knowledge will become prior and contribute to valuation.

## 2.2 Neurocognitive functions associated with decision-making

As illustrated in the overview above, decision-making is a complex cognitive process that engages a host of brain functions (Fig.1 right), from the traditional motivation (goal setting, needs, wants, self-regulation, Kuhl, 1986; Higgins. 2000; Doya, 2008), attention (selecting which pertinent information to acquire, Legrenzi et al., 1993; Endsley, 1995; Rolls, 2007; Summerfield and Egner, 2014), reasoning (planning the set of actions leading to goal completion, Azjen, 1991; Pennington & Hastie, 1993), value-expectancy (creating a mental model of the consequences of different alternatives, Sugrue et al., 2005; Hare et al., 2009), emotion (modulating motivation, Elster, 1998, simulating cost of different alternatives, Naqvi et al., 2006, estimating risk/benefit, integrating feedback into a learnt experience, Doya, 2008), and memory (retrieving priors/past contingencies between context and outcome, recalling beliefs, learning from feedback, Endsley, 1995; Rolls, 2007; Doya, 2008), to the more integrated functions described as situational awareness (a conscious experience of complex external and internal variables participating in decision making, Endsley, 1995) and habits (automatic behavior, Norman & Shallice, 1980; Ikosaka & Isoda, 2010; Redgrave et al., 2010 and biases, Bechara et al., 2000; De Martino et al., 2006).

## 2.3 Some factors modulating decision-making

Several internal and external circumstances modulate the decision-making process. Chiefly, uncertainty plays a large role at multiple levels (Hampshire & Hart, 1958; Doya, 2008). Uncertainty might take place on the side of information input: for external or internal reasons, information might be incomplete, inaccessible or noisy, gathered under temporal pressure (Johnston & Ratcliff, 2014), for instance when the rate of change in the world is faster than the rate of information accumulation (Brehmer, 1992), or with limited attentional resources; all to the effect that external context or internal state are insufficiently characterized; there might also be uncertainty on which information ("appropriate critical cues") to gather, -a matter of feedback quality which Endsley, (1995) found, beats feedback quantity (collect and process as much information as possible), because, in the latter case, human agent's capabilities are quickly overloaded-. Furthermore, the relationship between action and expected outcome might be unknown (novel situation, changing context): it might be that the causal link between undertaken action and actual outcome cannot be determined, weakening the future contingencies between them, or encouraging risk and exploratory behavior (Doya, 2008). Experimentally, uncertain outcome is embodied in the availability and readiness of feedback (Endsley, 1995). Without feedback in experimental tasks from game theory, people prefer large and uncertain outcomes (Jessup et al., 2008), that is, they are ready to take more risk (see also Doya, 2008). Feedback delays also negatively influence dynamic decision-making processes (when a series of decisions is needed in quick succession), e.g. in complex, dynamic decision-making tasks embedded in virtual microworlds (Brehmer, 1992), and in simulated inventory management (Diehl & Sterman, 1995). Another source of uncertainty in decision-making is that multiple goals might be co-present (Brehmer, 1992), which cannot all be satisfied simultaneously, and whose rankings are underdetermined (Kahneman & Tversky, 1986). In fact, some people have "pathologies" of decision-making described as "thematic vagabonding" wherein they constantly shift goals (Brehmer, 1992), which might be an extreme case of a more mundane manifestation of complex decision-making exerted in everyday circumstances. In all of those cases with growing uncertainty, decision-making is compromised, to the effect that optimal solutions are less frequently reached than in situations of certainty. And decision time, when freely available, is steeply increased.

A second important modulator of decision making is risk (Tversky and Kahneman, 1986), defined as the elevated likelihood that an action fails (risky action), or the likelihood that an outcome has devastating consequence on the agent (risky outcome, e.g. Doya, 2008). A body of work (Higgins, 2000; Crowe and Higgins, 1997) has demonstrated that depending on the level of risk, decision processes change from a style emphasizing promotion (maximize gains; minimize missed opportunities) to one emphasizing prevention (avoidance of bad outcomes, often realized by choosing no-action). Distinct theories (Edwards, 1954) and different computational and statistical models (Johnson & Ratcliff, 2014) have been proposed to deal with risky and riskless choices respectively.

Time is yet another factor that modulates decision-making. We have already introduced the effect of time pressure on information gathering (the notorious speed-accuracy tradeoff; Gold & Shalden, 2007; see exemplary model in Johnson & Ratcliff, 2014). In well-designed, experimental paradigms of dynamic decision-making (Brehmer, 1992), fast rates of change of information were found to be profoundly taxing for the agent making decision. Earlier, we had also reviewed feedback delays' effect on establishing the contingency between action selection and outcome. Time also plays a role in temporally unconstrained decision making: the agent might delay action until a confidence level is reached (Endlsey, 1995), a so-called "decision criterion" (e.g. Swets, 1996) that software systems, for good or bad, often lack as they are typically set on a fixed timetable to deliver decisions and actions. Feedback delay and more specifically delay-discounting (larger but delayed feedback requiring impulse control to maximize payoff) is yet another temporal factor with well-known consequences on decision-making mediated by motivational and emotional factors (Doya, 2008): people tend to accept a smaller but more immediate gain. The temporal structure of serial decision making also reveals some consecutive effects. Genovesio & Ferraina, (2014) reviewed experimental evidence demonstrating that previous trials affect performance (e.g. perceptual or motor conflicts that deteriorate current trial's performance due to the characteristics of previous trials; perceptual expectancy that enhances it). They identified separable contributions from previous goals and previous outcomes, and they proposed a neural architecture to support their maintenance in memory (see also Curtis & Lee, 2010).

And last but not least, emotions contribute enormous modulation to the decision-making process (Elster, 1991; Schwarz, 2000; Hastie & Dawes, 2001; Lowenstein & Lerner, 2003; Bechara, 2004; Naqvi et al, 2006); one of several reasons why earlier theories of the "rational agent" (an hypothetical human deemed by early economic research as always making choices in his/her best interest, see, e.g. Kahneman, 2002) was abandoned for more sophisticated models. In the next section, we first draw an overview of emotion theory before we examine its interplay with the several processes of decision making.

## 3 THEORIES OF AFFECTS AND EMOTIONS

There has not emerged an absolute consensus on a theory of human emotions (see outstanding debates in the volume edited by Ekman & Davidson, 1994; Oatley et al., 2006; Clore & Ortony, 2013), but in the following, we try to draw an operational framework that we envision to serve in future implementations of self-adaptive and co-adaptive software architectures where emotion would play a role. After providing a brief synthesis of the main theories of emotions, we examine the responses elicited in the brain and in the body by emotional episodes. We then identify the contributions of emotions to the processes of decision-making.

### 3.1 Taxonomy of Emotions and Temporality of Affects

Two longstanding debates in the field of emotion are (1) whether there exist qualitatively distinct emotions (Oatley & Johnson-Laird, 1990; Roseman et al., 1994; Ekman 1999; Lerner & Keltner, 2000; Plutchick, 2001; Lövheim, 2012 see also Nummenmaa et al., 2014) or whether emotions just exist on a continuum of valence (positive or negative) and intensity (Ekman, 1957, -a view later revised, see Ekman, 1999-; Russell, 1980; 2003; Ortony et al., 1988; Posner et al., 2008; Steunebrink et al., 2009; Clore & Ortony, 2013); and (2) for those acquiescing to the first proposition that discrete emotions do in fact exist at a fundamental (Ekman, 1999) or pragmatic level (Clore & Ortony, 2013), whether there are just a few basic (primary) emotions that combine to create the richness of emotional experience (e.g. smugness as happiness + contempt, see Ekman, 1999), or whether there are many different types of mutually-exclusive emotions (Ortony et al., 1988; Steunebrink et al., 2009), with their distinct autonomic and neural bases. Proponents of a model of discrete primary emotions which would be recombined into a rich fabric of secondary emotions include Ekman, Friesen & Ellsworth (1972/2013) with a model identifying 6 primary emotions as anger, disgust, fear, joy, sadness and surprise; to which another influential model by Pluchtik, (2001) adds two more: trust, and anticipation. There are other models using a much broader set of emotions, such as the 16 elements from the Geneva Emotion Wheel of Scherer, 2005; to the 22 prototypes of Ortony, Clore and Collins,

(1990; see also Steunebrink et al., 2009), notoriously recognized in the engineering literature and elsewhere as the OCC model.

Besides the type of emotion, there is also a question of how emotional experiences and affects organize themselves temporally (Fig.2). Emotions exist in the past, present and future (vertical axis of Fig.2). Thanks to memory processes, emotions that have happened in the past can be stored and recalled (see extensive work by Ledoux on fear conditioning, e.g. Ledoux, 1994; and Singer & Salovey, 1991; Schwarz, 2000; Baumeister et al., 2007). As exemplified in the description of decision-making processes, emotions can also be projected in the future, with the anticipatory mechanism of valuation that simulates the emotional consequences of prospective actions (Schwarz, 2000; Lerner & Keltner, 2000; Gross, 2002; Sugrue et al., 2005; Naqvi et al., 2006; Baumeister et al., 2007; D'Argembeau et al., 2008; Hare et al., 2009). Emotions and affects also occupy widely different durations (horizontal axis of Fig.2, and Frijda et al., 1991; Schwarz, 2000), from fractions of seconds (immediate emotions) to days and months (moods, defined as "prolonged core affect with no object" in Russell, (2003), see also Price, 2006; Singer & Savoley, 1991) and years (personality dispositions; Ledoux 1994; Lerner & Keltner, 2000; Plutchik, 2001; Anderson & Phelps, 2002; Scherer, 2005). The former (Fig.2 A-D) are task- or context-related; the latter (Fig.2 E-F) task-unrelated, although we'll later see that they blend in the psychophysiological response and make direct contribution to ongoing behavior.

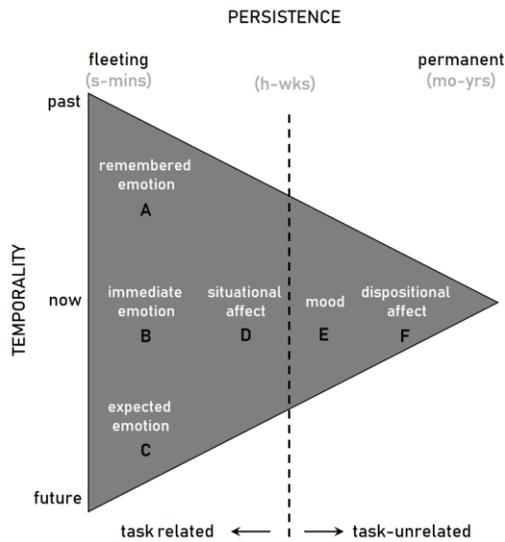

Figure 2: organization of emotions according to temporal occurrence (vertical axis) and persistence (horizontal)

## 3.2 Brain processes and body-wide responses

This previous question of temporality also entails a neurocognitive dimension, that is, the sequence of mental processes that gives rise to affective states. Before we examine the processes themselves, we need to expose the two classes of experienced emotions that are distinguished in many theoretical frameworks: automatic Vs. extended/full-blown (Ekman, 1999; Baumeister et al., 2007, see also Lerner & Keltner, 2000; Gold & Shalden, 2007; Zheng et al., 2007). Note that this dichotomy between automatic and extended emotional processing partially but incompletely overlaps with a distinction between pre-conscious or conscious processes. For instance, the affects labelled D,F in Fig.2 are typically unconscious. Some have proposed to reserve the terminology of 'mood' and 'emotions' for the slow and extended process, whereas 'affect' would be used for the automatic process (Baumeister et al., 2007). The dichotomy between preconscious and conscious emotions is challenged by another arm of the emotional research field, which contends that appraisal (Roseman & Evdokas, 2004) or interoception (Critchley & Harrison, 2013), -which can be conceived as a conscious cognitive effort to perceive one's own internal states-, causes emotion (Damasio et al., 2000; Prinz, 2004; Bechara, 2004; Roseman & Evdokas, 2004; Naqvi et al., 2006; Critchley & Harrison, 2013).

A meta-analysis of models of emotional processes that accounts for all those distinctions is proposed by Scherer (2009), with a valuable summary figure 1 that our readers are referred to. Across the four models that are compared, appraisal, arousal and emotional experience are related through different causal pathways depending on their originating theory. Insisting on the evolutionary adaptive nature of emotion (emotions serve to set the organism in movement, fight or flight), Fridja (2009) defines the key elements as appraisal, affect, action readiness, autonomic arousal and expressive behavior (a mechanism for social communication). Through his study of emotional regulation (an important adaptive process that prevents emotions to wreak havoc on the organism's function), Gross (2002) starts his model with situation selection (akin to motivated behavior); followed by attention, cognition and reappraisal (a process for downregulation of emotion) leading to a new behavioral response. Unsurprisingly when one considers the overlap of emotional and cognitive systems in the brain (e.g. Gray, 1990), those models of emotional processes often resemble the decision making's, as previously described in section 2.1.

Emotions also affect the body via the organism's autonomic and neuro-endocrine systems (Ledoux, 1993; Thayer & Lane 2000; Russell, 2003; Anders et al., 2004; Dagleish, 2004; Scherer, 2005; Kreibig, 2010; Critchley & Harrison, 2013). The main peripheral consequences of interest to us are cardiovascular (heart rate, vasodilatation), thermal (change in body temperature), pupillary (change in pupil diameter), respiratory (e.g. faster breathing), galvanic (electrodermal change in skin conductance property due to sweat glands), and of course, behavioral (motor reactions, including, e.g. changes in body posture, voice pitch, flushing, blushing, smiling, grinning, laughing, frowning, weeping and crying, Elster, 1998). A detailed review of specific changes elicited in distinct emotions can be found in Kreibig, (2010), see also Levenson 2004). All of those neurophysiological domains are targets for techniques aimed at providing a quantitative description of emotion; as are brain responses (see section 5).

### 3.3 Emotional influences on decision making

Emotions and decision-making are profoundly intertwined, as noted in Schwarz in 2000, with emotion affecting decision-making, and decision-making impacting the person's feelings. How do human passions help or hurt decision-making? Neurology has indicated for quite some time that emotions, -which were initially perceived as detrimental to decision-making in behavioral economics-, in fact do serve a fundamental guiding purpose. In seminal work, Damasio and his team have shown that patients carrying lesions in key brain regions that suppress emotions also find themselves incapable of making sound decisions (e.g. Damasio et al., 2000; Bechara, 2004; Naqvi et al., 2005) see also Tversky and Kahneman, 1986; Hastie & Dawes, 2001; Picard, 1997; Elster, 1998). Nonetheless, extreme and poorly regulated emotions are known to harm decision-making, cognition and behavior (Crichtley & Harrison, 2013; Baumeister et al., 2007), by turning attention to internal states at the expense of thoughts, feelings, sensory and cognitive processes (Elster, 1998), which is why some mechanisms of emotional self-regulation and reappraisal are in place in organisms (Crowe and Higgins, 1997; Thayer & Lane, 2000; Ochsner et al., 2002; Gross, 2002).

A primary role that emotions play in decision-making is through motivational behavior (Roseman, 1984; Critchley & Harrisson, 2013): emotions are rewarding, therefore, they are a useful control system for the reward-seeking organisms (Gold and Shalden, 2007) to change action readiness (Frijda, 2004), to set behavior in motion when the emotional outcome is pursued or to the contrary, to prevent it when outcome should be avoided (Baumeister et al., 2007), to "unprocrastinate" (Elster, 1998).

By modifying arousal through its autonomic component, emotion also modifies information processing and attention (Anderson & Phelps, 2002; Loewenstein & Lerner, 2003; Prinz, 2004) for instance, by increasing bottom-up decision-making and attention to details when mood is negative and indicative of a challenging environment (Schwarz, 2000); by shining attention to dangers and risks (Prinz, 2004) or by creating perceptual expectancy of reward (Lauwereyns et al., 2002). On a counterpart, positive moods also have the propensity to elicit a decision-making process that ignores environmental cues and relies on pre-existing knowledge (Schwarz, 2000) and relatedly, emotions bias the perception of new data toward initial cognitive expectations (Lerner & Keltner, 2000).

A crucial component of emotion is through the valuation system that predicts the expected outcome of different alternatives in the decision-making process (Schwarz, 2000; Loewenstein & Lerner, 2003; Navqi et al., 2006; Baumeister et al., 2007; d'Argembeau et al., 2008; Hare et al., 2009; Crichtley & Harrison, 2013), -and many a theory posits that this valuation is accomplished by a fully embodied simulation of future emotions (Damasio et al., 2000; Bechara, 2004; Naqvi et al., 2005; Clore and Ortony, 2013)- (with the convenience that those signals can be detected in brain and behavior, see thereafter).

Continuing along the process of decision-making, experienced emotion also contributes to reinforcement learning (Elster, 1998; Russell, 2003; Sugrue et al., 2005; Baumeister et al., 2007; Verschure et al., 2014), increases memory (Ledoux, 1994; Baumeister et al., 2007) of both positive and negative emotions and facilitates recall (Anderson & Phelps 2002), especially when the affective context/tone at recall is congruent with that from encoding (Singer & Salovey, 1988). However, other processes of emotional regulation, such as emotional suppression, have been found to impair memory (whereas others, like reappraisal, leave it intact (Gross, 2002). In reviewing some experimental work, Schwarz (2000) also notes that the dynamics of emotional memory is not consistent across time, but that two periods are privileged: that of peak affect and that ending the emotional experience. We can predict that those periods will make distinct contributions to subsequent decision making through the valuation system.

Finally, it should not be forgotten that emotion carry a communicative function between individuals. In that respect, emotions impact individual decision-making made in social context, as well as social decision-making (Elster, 1998; Loewenstein et al., 1989; Van Kleef et al., 2011, see also Critchley & Harrison 2013; Dunn & Schweitzer, 2005). Studying interpersonal decision-making Loewenstein et al., (1989) implicate the nature of the relationship (e.g. friends or enemy) and the type of problem (competition, cooperation) as key factors, and they underline the fact that interpersonal contexts likely involve larger emotions, thereby magnifying the above-described effects of emotions in individual decision-making. The comprehensive experimental review by Van Kleef et al., 2011 reveals that the direction of affect (self-other and its reciprocal, or mutual affects) leads to an array of distinct influences on decision-making. For instance, others' emotions can become a source of information used to guide one's own decisions (risk and valuation outcome), or emotional contagion can have collective effects on a group's shared-task. Self-emotions in social contexts are also important motivational sources (encouraging the social regulation of action tendencies, Elster 1998; driving competition and cooperation, Van Kleef et al., 2011; see also Immordino-Yang & Damasio, 2007 for an educational account). Some emotions (e.g. happiness, distress) facilitate reciprocal or unilateral cooperation and therefore improve (collective) decision-making (but also triggers competition in adversarial situations), whereas other emotions (e.g. anger) are detrimental to cooperation (Van Kleef et al., 2011). It should also be noted that eliciting strong emotions in others may be of tactical advantage in competitive contexts, by exploiting the disorganization of decision-making that is likely to arise in the receiver (as is commonly practiced in sport and military contexts).

To summarize (Table 1), current emotions (Fig.2B) as well as moods, situational and dispositional affects (Fig.2-D-F) alter motivational components of decision-making, attentional and perceptual, and memory encoding processes. Future emotions (Fig.2C) play a role in guiding the motivational system, but their key role is in the valuation system, which uses a simulation of affect to select a course of action that is beneficial for the individual. Past emotions (Fig.2.A) feed information into this valuation system by virtue of the mechanism of memory recall (embodied simulation), but also alter perceptual processing, attention, and response selection (e.g. habits). It emerges from this temporally complex view (see also Loewenstein & Lerner, 2003 and Schwarz, 2000 for excellent temporal perspectives on emotion and decision-making) that the challenges for self- and co-adaptive systems feeding from human-emotions-in-the-loop will be, -not only to provide accurate and valid measures of emotional states-, but also to disentangle the several processual contributions of emotions into the decision-making process, a matter that a careful dynamical perspective might hope to resolve.

Table 1: emotional factors influencing decision-making

| Emotion | Effect on decision-making | /Fig.2 |
|---|---|---|
| Current emotions | motivational, perceptual, attentional, memory encoding processes | A |
| Past emotions | valuation system, perceptual processing, attention, response selection | B |
| Future emotions | motivational and valuation systems | C |
| Situational/disposi-tional affects, moods | motivation, attention, perception and memory | D-F |

## 4 HUMAN-INSPIRED ADAPTATION IN SOFTWARE SYSTEMS

Decision-making is central to software with self- and co-adaptive architectures, which typically have an adaptive decision engine for their autonomous function and another ("co-adaptive") decision engine to determine explicit inclusion of human-in-the-loop from a background of non-obtrusively collected environmental and psycho-physiological variables (Lloyd et al., 2017). In many conventional and self-adaptive software systems and in robots, an historical and still frequent approach to decision-making is with decision trees and associated 'if-then' rules. Those sets of rules can run deep with increasingly complex software, but they remain deterministic models for decision-making, and they are at odd with what biology and human behavior have had to offer after hundreds of millions of years chance-experiment with evolution (Ni et al., 2016). Due to their rigidity, deterministic rules are at risk of settling in local minima that miss the true (deeper) optimal state of the system (e.g. a learning system for children with autism, which fails to explore new educational contents without overwhelming the fragile trust and self-confidence of the learner, unless a stochastic component is implemented into the task switching program, Petersen et al., unpublished). Deterministic rules also do not perform particularly-well in changing contexts, when there is uncertainty in the information input or in the user's behavior, which is why there currently is a movement in the software community to develop systems that follow a less predetermined behavior, including systems inspired from human emotions and cognition (e.g. De Lemos, 2015). Other theoretical or practical solutions to the rigidity of fixed rules should be mentioned and include stochastic optimization, fuzzy logic, simulated annealing, cross-entropy, chaos and metastability (Lee, 1990; Pai & Hong, 2006; Moreno et al., 2017; Nara, 2003; Li et al., 2008; Tognoli & Kelso, 2014; see also Harman, 2007).

Because emotions increase complexity and entropy, while at the same time demonstrating adaptive behavior in biological systems (Picard, 1995; Elster, 1998; Damasio et al., 2000; Hastie & Dawes, 2001; Loewenstein & Lerner, 2003; Bechara 2004; Naqvi et al., 2006; Baumeister et al., 2007), they are an ideal substratum upon which to build self- and co-adaptive software systems. In them, emotional events are nudged to move to new states. We envision scenarii (1) where emotions affect the software decision-making in exactly the same ways as they operate in human, so as to explore human-inspired self-adaptive software, discover their strength and weaknesses, and refine a co-adaptive model where human and machine operate to the maximal benefit of both; and (2) to use human emotional inputs in new ways. For instance, to palliate to a decision-making module lacking situational awareness of changing contexts, humans could become sensors of uncertainty, and consequently modulate the decision criterion that the software would employ.

## 5 BUILDING HUMAN EMOTIONS IN THE LOOP

Machines with emotional intelligence is one of the key ideas of affective computing where machines (software systems) are given the abilities to sense human's emotions, feelings, perceptions and cognitive states [Picard 1997, 2003]. Emotion is a fundamental aspect of human experience, spanning for example active learning, social life and rational decision making. In co-adaptive software systems, where human and machine form a very close symbiotic partnership, the desire to design software systems capable of recognizing human's emotion, behaviors and cognitive states is essential. The ultimate goal of such co-adaptive systems is to "feel" and "anticipate" human's emotion and cognitive states, so that computer systems can react proactively [Huang & Tognoli 2014, Lloyd, Huang & Tognoli 2017], and vice versa, can include in the loop the rational decisions that humans make with the support of their emotional states.

Most human emotion recognition approaches are based on Ekman's six or seven basic facial expressions –anger, content, fear, disgust, happiness, sadness, and surprise [Ekman 2003; Ekman et al., 2013] and use image or movies of face as input material. From the technical perspective, the advancement of machine learning algorithms and artificial intelligence makes the emotion recognition software more accurate and faster. One of the dominant open software tools in face recognition is Open Source Computer Vision Library (OpenCV) [OpenCV]. Its object detection uses Haar feature-based cascade classifiers and can achieve high speed recognition. Another good option of open source software is Dlib face detection [DLib, Dlib blog]. It is built in C++ and has the ability to recognize face landmarks. Its shape predictor can be used to recognize difference emotions, is high accurate and much more detailed.

There are some face datasets that can be used to train emotion recognition software systems. For example, the Cohn-Kanada AU-Coded facial expression database contains nearly 100 subjects [Kanade, Cohn, & Tian 2000]. The Japanese Female Facial Expression (JAFFE) database contains 60 Japanese subjects, each provides seven facial expressions (6 basic facial expressions + 1 neutral). [Lyons, Akamatsu etc 1998, JAFFE database], see also the Karolinska Directed Emotional Faces (KDFS) [Lundqvist, Flykt, and Öhman 1998].

Besides open source software, there are quite a few readily available software packages for emotion recognition. One of the most representative software systems is Affectiva [Affectiva]. Affectiva uses unobtrusively measures unfiltered and unbiased facial expression of emotions to measure seven human emotion metrics anger, contempt, disgust, fear, joy, sadness and surprise. Affectiva provides SDK and API for end users to customize their unique emotion recognition needs. Because humans use a set of non-verbal cues to expression their emotion, Affectiva now is adding speech capability to their systems (beta test). Note that multimodal emotion recognition (emotion recognition conducted in parallel with multiple techniques) increasingly emerges as an important strategy for accurate quantification of emotions.

FaceReader i[FaceReader] is another commercial software systems that is used on usability studies, psychology, education and market etc. The drawback of FaceReader is that it doesn't provide SDK and API for customization.

Moving away from human's overt emotion recognition (facial expressions, behavioral and cognitive states), recently, at Consumer Electronics Show (CES) in Las Vegas, imec introduced a prototype of an EEG headset for emotion Detection. This wearable EEG monitoring system can measure emotions and cognitive processes in the brain [EEG Headset for Emotion Detection 2018]. This is a major breakthrough from traditional EEG brain scan on diagnosing medical conditions (epilepsy or sleep disorders) to detect emotions beyond the medical field. It opens the door for applications, such as e-Learning, and emotion-based e-gaming.

Advancement of AI has led to breakthroughs in humanoid robots' development. Pepper [Pepper], the humanoid emotional robot owned by Softbank, designed with the ability to read emotions. Rather than focusing on domestic use as most robots do, Pepper is intended to engage people, make them happy and enhance their life. Sophia developed by Hanson Robotics is another exemplar of how human interacting with technology. Sophia can mimic human facial expressions, gestures and can converse on s variety of topics. Those humanoid robot developments have changed the way human interact with software systems and technology. Finally, at the research frontiers are a variety of techniques that target the diverse autonomic and neural manifestations of emotions. Although they require ad-hoc hardware and custom-build software, they should be recognized as potential direction for future engineering design. They include Heart Rate Variability (HRV) and vascular tone; temperature changes, change in pupillary diameters, respiratory measures, electrodermal responses (e.g. Zhang et al., 2016) and the examination of postural dynamics.

## 6 CONCLUSIONS AND PERSPECTIVE

Inspired by those two reviews of the psycho-physiological processes of emotion and decision-making, we present the conceptual blueprint of a Human-Computer Interaction system whose software has both self- and co-adaptive capabilities, and that uses human-inspired decision-making process and emotions (Fig.3). The software's decision-making module (blue, top left) is not a system with "if-then" statements or decision trees that is typical of conventional software systems. Its architecture mirrors human decision-making's (blue, bottom left). In its self-adaptive mode, the software aims to function autonomously without human intervention (see also, e.g. Lloyd et al., 2017 for exemplary system). It is hoped that mimicking of human decision-making processes will augment the autonomy and adaptability of the software, an hypothesis that remains to be tested in a phase of experimental software application.

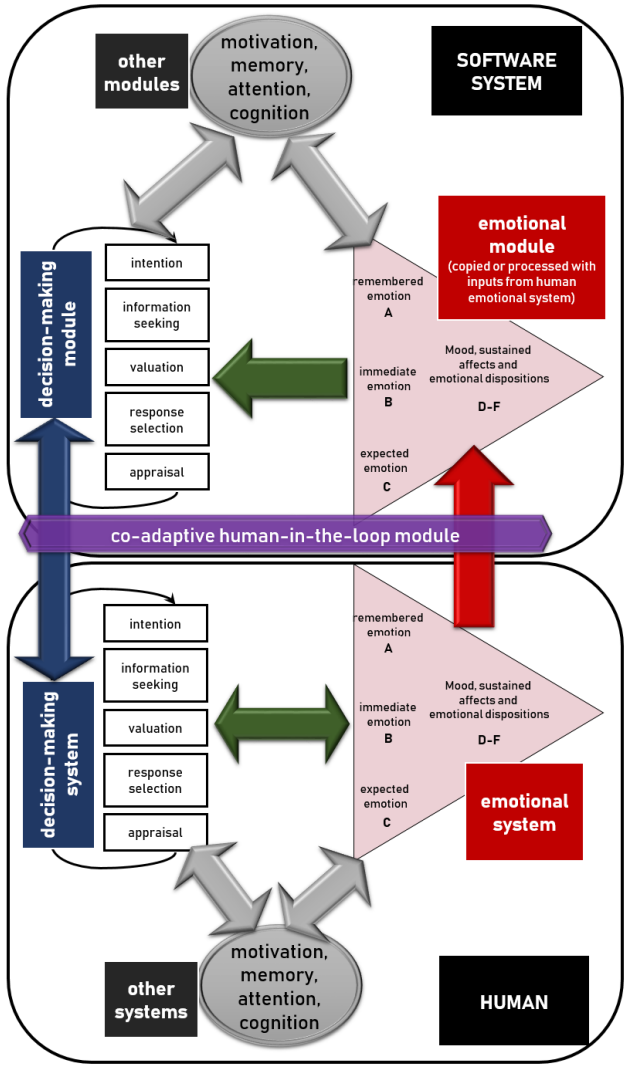

Fig.3: A Human Computer Interaction system with self-adaptive and co-adaptive capabilities, which includes human-inspired decision-making and human-sourced emotions. See details in text.

The autonomous capability of such system does not mean however, that the software lacks input from humans. Especially, in the area of decision-making, computers are known to lack emotional wisdom and gut's feelings (De Lemos, 2015; see also Picard, 1995), an opportunity for human emotions to help. Until a later time when affective computers are realized (Picard, 1995) and following the strategy envisioned in Lloyd et al., (2017), human emotions (bottom right, red; which are estimated with any suitable combination of techniques reviewed in Section 5) can be sourced unobtrusively to the software system (top right, red) either as pure copy or with some processing (e.g. if emotions are monitored from multiple humans or require deconvolution…). The software's emotional module (top right, red) then interacts with the software's decision-making module in charge of core tasks (blue, top left) via a unidirectional connection (green arrow, top).

The second component of this Human-Computer Interaction system is the co-adaptive module (purple, center), which mediates obtrusive, intentional, cooperative interactions between human and software for the realization of shared-tasks (e.g. consensus-forming tasks or tasks exploiting complementary skills of humans and machines). In earlier work (Lloyd et al., 2017), interaction was mediated by a psychometric model of task performance, the OCW model (Blumberg & Pringle, 1982; Camara et al., 2015), monitored unobtrusively with EEG. The present conceptual blueprint retains those features and enhances them with shared emotional and decision-making space. (vertical arrows).

In concluding, this project inscribes itself in line with the prophetic paper by Licklider [1950] of a man-computer symbiosis. By taking inspiration from human decision making and human emotion in individual and social contexts, we anticipate a smoother interaction between computers and people, and progress in their adaptability and response to uncertainty.